\documentclass{llncs}
\usepackage{amssymb}
\usepackage{amsmath}
\def\F{\mathbb{F}}
\def\Z{\mathbb{Z}}
\def\simplex{\delta}
\def\Simplex{\Delta}
\def\RLife{R_{\textbf{\emph{Life}}}^\simplex}
\def\PLife{P_{\emph{\textbf{Life}}}}
\newcommand{\esymm}[2]
{
\sigma_{#1}\left(#2\right)
}
\def\blcell{\makebox(0,0){\rule{2pt}{2pt}}}
\newcommand{\Wpqrsbare}[2]
{
\begin{picture}(32,24)(#1,#2)
\multiput(5,17)(14,0){3}{\circle{8}}
\put(5.2,16.8){\makebox(0,0){\scriptsize$p$}}
\put(18.9,16.8){\makebox(0,0){\scriptsize$q$}}
\put(33.2,16.8){\makebox(0,0){\scriptsize$r$}}
\put(19.2,0){\makebox(0,0){\scriptsize$s$}}
\put(19,0){\circle{8}}
\put(16.7,3.1){\line(-1,1){10}}
\put(19,4){\line(0,1){9}}
\put(21.3,3.1){\line(1,1){10}}
\end{picture}
}
\newcommand{\Wprsbare}[2]
{
\begin{picture}(32,24)(#1,#2)
\multiput(5,17)(28,0){2}{\circle{8}}
\put(5.2,16.8){\makebox(0,0){\scriptsize$p$}}
\put(33.2,16.8){\makebox(0,0){\scriptsize$r$}}
\put(19.2,0){\makebox(0,0){\scriptsize$s$}}
\put(19,0){\circle{8}}
\put(16.7,3.1){\line(-1,1){10}}
\put(21.3,3.1){\line(1,1){10}}
\end{picture}
}
\newcommand{\Wpsbare}[2]
{
\begin{picture}(20,20)(#1,#2)
\put(0.6,15.6){\circle{8}}
\put(1,15.5){\makebox(0,0){\scriptsize$p$}}
\put(16.2,0){\makebox(0,0){\scriptsize$s$}}
\put(16,0){\circle{8}}
\put(13,2.4){\line(-1,1){10}}
\end{picture}
}
\newcommand{\Wqsbare}[2]
{
\begin{picture}(6,20)(#1,#2)
\put(1,15.6){\circle{8}}
\put(0.9,15.5){\makebox(0,0){\scriptsize$q$}}
\put(1.2,0){\makebox(0,0){\scriptsize$s$}}
\put(1,0){\circle{8}}
\put(1,4){\line(0,1){7.5}}
\end{picture}
}
\newcommand{\Wrsbare}[2]
{
\begin{picture}(20,20)(#1,#2)
\put(17.1,15.6){\circle{8}}
\put(17.2,15.5){\makebox(0,0){\scriptsize$r$}}
\put(2.2,0){\makebox(0,0){\scriptsize$s$}}
\put(2,0){\circle{8}}
\put(5,2.4){\line(1,1){10}}
\end{picture}
}
\newcommand{\Wpqsbare}[2]
{
\begin{picture}(32,24)(#1,#2)
\multiput(5,17)(14,0){2}{\circle{8}}
\put(5.2,16.8){\makebox(0,0){\scriptsize$p$}}
\put(18.9,16.8){\makebox(0,0){\scriptsize$q$}}
\put(19.2,0){\makebox(0,0){\scriptsize$s$}}
\put(19,0){\circle{8}}
\put(16.7,3.1){\line(-1,1){10}}
\put(19,4){\line(0,1){9}}
\end{picture}
}
\newcommand{\Wqrsbare}[2]
{
\begin{picture}(18,24)(#1,#2)
\multiput(5,17)(14,0){2}{\circle{8}}
\put(4.9,16.8){\makebox(0,0){\scriptsize$q$}}
\put(19.2,16.8){\makebox(0,0){\scriptsize$r$}}
\put(5.2,0){\makebox(0,0){\scriptsize$s$}}
\put(5,0){\circle{8}}
\put(5,4){\line(0,1){9}}
\put(7.3,3.1){\line(1,1){10}}
\end{picture}
}
\newcommand{\set}[1]{\left\{#1\right\}} % Set of elements

\newcommand{\putxi}[3]
{
\put(#1,#2){\makebox(0,0){\circle{15}}}
\put(#1,#2){\makebox(0,0){$\hspace*{-12pt}x_#3$}}
}

\begin{document}
\title{On Compatibility of Discrete Relations}
\titlerunning{Discrete Relations}
\author{Vladimir V. Kornyak}
\institute{Laboratory of Information Technologies \\
           Joint Institute for Nuclear Research \\
           141980 Dubna, Russia \\
           \email{kornyak@jinr.ru}}
\authorrunning{Vladimir V. Kornyak}
\maketitle
\begin{abstract}
An approach to compatibility analysis
of systems of discrete relations is proposed.
Unlike the Gr\"obner basis technique, the proposed scheme
is not based on the polynomial ring structure.
It uses more primitive set-theoretic and topological
concepts and constructions. We illustrate the approach
by application to some
two-state cellular automata. In the two-state case the
Gr\"obner basis method is also applicable,
and we compare both approaches.
\end{abstract}

\section{Introduction}
A typical example of a system of discrete relations
is a cellular automaton.
Cellular automata are used successfully
in a large number of applications.%
\footnote{Comparing expressiveness of cellular automata and
differential equations, T. Toffoli writes \cite{Toffoli}: ``Today,
it is clear that we can do all that differential equations can do,
and more, because it is differential equations that are the poor
man's cellular automata --- not the other way around!''}
Furthermore, the concept of cellular automaton can be generalized,
and we consider the following extension of the standard notion of
a cellular automaton:
\begin{enumerate}
    \item
Instead of regular uniform lattice representing
the space and time in a cellular automaton,
we consider more general
\emph{abstract simplicial complex}
$K = \left(X,\Simplex\right)$
(see, e.g., \cite{HiltonWylie}).
Here $X = \set{x_0,x_1,\ldots}$ is a finite (or
countably infinite)
 set of \emph{points};
$\Simplex$ is a collection of subsets of $X$
such that (a)~ for all $x_i\in X,~\set{x_i}\in\Simplex$;~
(b)~ if $\tau\subseteq\simplex\in\Simplex$,~ then $\tau\in\Simplex$.\\
The sets $\set{x_i}$ are called \emph{vertices}.
We say $\simplex\in\Simplex$ is  a $k-$\emph{simplex}
of \emph{dimension} $k$ if
$|\simplex|=k+1,$~ i.e., ~$\dim \simplex =|\simplex|-1.$
The \emph{dimension of complex} $K$ is defined as
the maximum dimension of its constituent simplices~
$\dim K = \max\limits_{\simplex\in\Simplex} ~\dim \simplex.$\\
If $\tau\subseteq\simplex,~\tau$ is called a \emph{face} of $\simplex$.
Since any face of a simplex is also a simplex, the topological structure
of the complex $K$, i.e., the set $\Simplex$ is uniquely determined
by the set of \emph{maximal simplices} under inclusion.

One of the advantages of simplicial complexes over regular lattices
is their applicability to models with dynamically emerging and
evolving rather than pre-existing space-time structure.
    \item
The dynamics of a cellular automaton is determined by a
\emph{local rule}
\begin{equation}
    x_{i_k}=f\left(x_{i_0},\ldots,x_{i_{k-1}}\right).
    \label{localrule}
\end{equation}
In this formula $x_{i_0},\ldots,x_{i_k}\in X$ are interpreted
as discrete variables taking values in a finite set of states
$S$ canonically represented as
$$S = \set{0,\ldots,q-1}.$$
The set of points $\set{x_{i_0},\ldots,x_{i_{k-1}}}$
is called the \emph{neighborhood}.
The point $x_{i_k}$ is considered as the ``next time step''
match of some point, say $x_{i_{k-1}}$, from the neighborhood.

A natural generalization
is to replace function (\ref{localrule}) by a \emph{relation}
on the set $\set{x_{i_0},\ldots,x_{i_k}}.$
In this context, local rule (\ref{localrule}) is a
special case of relation. Relations like
  (\ref{localrule}) are called
\emph{functional relations.}
They are too restrictive in many applications.
In particular, they violate in most cases the symmetry
among points $x_{i_0},\ldots,x_{i_k}.$
Furthermore, we will see below that the functional
relations, as a rule, have non-functional consequences.
\end{enumerate}
We can formulate some natural problems concerning the above
structures:
\begin{enumerate}
    \item
\emph{Construction of consequences.} Given a relation
$R^\simplex$ on a
set of points $\simplex$, construct non-trivial relations
$R^\tau$ on subsets $\tau\subseteq\simplex$, such that
$R^\simplex\Rightarrow R^\tau.$
    \item
\emph{Extension of relation.}
Given a relation $R^\tau$ on a subset $\tau\subseteq\simplex$,
extend it to relation $R^\simplex$ on the superset $\simplex$.
    \item
\emph{Decomposition of relation.} Given a relation
$R^\simplex$ on a set $\simplex,$
decompose $R^\simplex$ into combination of relations
on subsets of $\simplex.$
    \item
\emph{Compatibility problem.}   Given a collection of relations
$\set{R^{\simplex_1},\ldots,R^{\simplex_n}}$ defined on sets
$\set{\simplex_1,\ldots,\simplex_n}$, construct
relation $R^{\cup_{i=1}^{n}\simplex_i}$  on the union
$\bigcup_{i=1}^{n}\simplex_i$, such that
$R^{\cup_{i=1}^{n}\simplex_i}$ is compatible with
the initial relations.
    \item
\emph{Imposing topological structure.} Given a relation
$R^X$ on a set $X$, endow  $X$ with a structure of
simplicial complex consistent with the
decomposition of  the relation.
\end{enumerate}
If the number of states is a power of a prime, i.e., $q=p^n$, we can always%
\footnote{Due to the functional completeness
of polynomials over $\F_q$ (see \cite{Lidl}) any function mapping
$k$ elements of $\F_q$ into $\F_q$ can be realized by a polynomial.}
represent any relation over $k$ points
$\set{x_1,\ldots,x_k}$ by the set of zeros of some
polynomial from the ring
 $\F_q\left[x_1,\ldots,x_k\right]$ and study the compatibility
problem by the standard Gr\"obner basis methods.
It would be instructive to look at the compatibility problem from the
set-theoretic point of view cleared of the ring structure influence.

An example from fundamental physics is the \emph{holographic
principle} proposed by G. 't Hooft and developed by many authors
(see  \cite{Hooft00,Bousso}). According to 't Hooft the
combination of quantum mechanics and gravity implies that the
world at the Planck scale can be described by a three-dimensional
discrete lattice theory with a spacing of  the Planck length
order. Moreover, a full description of events on the
three-dimensional lattice can be derived from a set of Boolean
data (one bit per Planck area) on a two-dimensional lattice at the
spatial (evolving with time) boundaries of the world. The transfer
of data from two to three dimensions is performed in accordance
with some local relations (constraints or laws) defined on
plaquettes of the lattice. Since the data on points of the
three-dimensional lattice are overdetermined, the control of
compatibility of relations is necessary. Large number of
constraints compared to the freedom one has in constructing models
is one of the reasons why  no completely consistent mathematical
models describing physics at the Planck scale have been found so
far.

\section{Basic Definitions and Constructions}
The definition of \emph{abstract} $k$-simplex as a set of $k+1$
points is motivated by the fact that $k+1$ points generically
embedded in Euclidean space of sufficiently high dimension
determine $k$-dimensional convex polyhedron. The abstract
combinatorial topology only cares about how the simplices are
connected,
and not how they can be placed within whatever spaces.%
\footnote{There are mathematical structures of non-geometric
origin, like \emph{hypergraphs} or \emph{block designs}, closely
related conceptually to the abstract simplicial complexes.} We
need to consider  also $k$-point sets which we call
$k$-\emph{sets}. Notice that $k$-sets may or may not be
$(k-1)$-simplices.

A relation is defined as a subset of a Cartesian
product $S\times\cdots\times S$ of the set of states.
Dealing with the system of relations determined over
different sets of points we should indicate the
correspondence between points and dimensions of
the hypercube $S\times\cdots\times S.$
The notation $S^{\set{x_i}}$ specifies the set $S$
as a set of values for the point $x_i.$ For the $k$-set
$\simplex=\set{x_1,\ldots,x_k}$ we denote
$S^\simplex \equiv S^{\set{x_1}}\times\cdots\times
S^{\set{x_k}}.$

 A \emph{\textbf{relation}} $R^\simplex$ over a $k$-set
 $\simplex=\set{x_1,\ldots,x_k}$ is any subset
 of the hypercube $S^\simplex$, i.e.,
  $R^\simplex\subseteq S^\simplex.$
We call the set $\simplex$ \emph{domain} of the relation
$R^\simplex$.
The relations $\emptyset^\simplex$ and $S^\simplex$ are called
\emph{empty} and  \emph{trivial}, respectively.

Given a set of points $\simplex$,
its subset $\tau\subseteq\simplex$ and relation $R^\tau$
over the subset $\tau$,
we define \emph{\textbf{extension}} of $R^\tau$ as the relation
$$R^\simplex=R^\tau\times S^{\simplex\setminus\tau}.$$
The procedure of extension allows one to extend relations
$R^{\simplex_1},\ldots,R^{\simplex_m}$ defined on different
domains to the common domain, i.e., the union
$\simplex_1\cup\cdots\cup\simplex_m$.

Now we can construct the \emph{\textbf{compatibility condition}}
of the system of relations $R^{\simplex_1},\ldots,R^{\simplex_m}.$
Naturally this is intersection
of extensions of the relations to the common domain
$$
R^\simplex = \bigcap_{i=1}^m \left(R^{\simplex_i}\times
S^{\simplex\setminus\simplex_i}\right),~ \text{where}~~
\simplex=\bigcup_{i=1}^m\simplex_i.
$$
We call the compatibility condition $R^\simplex$
the \emph{\textbf{base relation}} of the system of relations
$R^{\simplex_1},\ldots,R^{\simplex_m}.$
If the base relation  is
empty, the relations
$R^{\simplex_1},\ldots,R^{\simplex_m}$
are \emph{incompatible.}
Note that in the case $q=p^n$ the compatibility
condition can be represented by a single polynomial,
in contrast to the Gr\"obner basis approach
(of course, the main aim of the Gr\"obner basis computation
--- construction of basis of polynomial ideal ---
is out of the question).

A relation $Q^\simplex$ is a \emph{consequence} of relation
$R^\simplex$,~ if $R^\simplex\subseteq Q^\simplex\subseteq
S^\simplex,$~ i.e.,~ $Q^\simplex$ is any superset
of $R^\simplex.$
Any relation can be represented in many ways by
intersections of different sets of its consequences:
$$
R^\simplex = Q^{\tau_1}\cap\cdots\cap Q^{\tau_r}.
$$
We call such representations \emph{decompositions}.

In the polynomial case $q=p^n$, any possible Gr\"obner
basis of polynomials representing the relations
$R^{\simplex_1},\ldots,R^{\simplex_m}$
corresponds to some decomposition
of the base relation $R^\simplex$ of the system
$R^{\simplex_1},\ldots,R^{\simplex_m}$.
However, the decomposition implied by a Gr\"obner
basis may look accidental from our point of view and
if $q\neq p^n$ such decomposition
is impossible at all.

The total number of all consequences (including
$R^\simplex$ itself and the trivial relation $S^\simplex$)
is, obviously, $$2^{\left(q^k-|R^\simplex|\right)}.$$
In our context it is natural to distinguish the consequences
which are reduced to relations over smaller sets of points.

A nontrivial relation $Q^\tau$ is called
\emph{\textbf{proper consequence}} of
relation $R^\simplex$ if
$\tau$ is a proper subset of $\simplex$, i.e.,
$\tau\subset\simplex$, and relation
$Q^\tau\times S^{\simplex\setminus\tau}$
is consequence of $R^\simplex$.

There are relations without proper consequences and these
relations are most fundamental for a given number of points
$k$. We call such relations \emph{\textbf{prime}}.

If relation $R^\simplex$ has proper consequences
$R^{\simplex_1},\ldots,R^{\simplex_m}$
we can construct its \emph{\textbf{canonical decomposition}}
\begin{equation}
R^\simplex = PR^\simplex\bigcap\left(\bigcap_{i=1}^m \left(R^{\simplex_i}\times S^{\simplex\setminus\simplex_i}\right)\right),
\label{canondec}
\end{equation}
where the factor
$PR^\simplex$, which we call the \emph{\textbf{principal factor}},
is  defined as
$$
PR^\simplex=R^\simplex\bigcup\left(S^\simplex\setminus
\bigcap_{i=1}^m \left(R^{\simplex_i}\times S^{\simplex\setminus\simplex_i}\right)
\right).
$$
The principal factor is the relation of maximum ``freedom'',
i.e., closest to the trivial relation but
sufficient to restore $R^\simplex$ in combination with the proper consequences.

If the principal factor in canonical decomposition (\ref{canondec})
is trivial, then  $R^\simplex$
can be
fully reduced to relations over smaller sets of points.
We call a relation $R^\simplex$ \emph{\textbf{reducible}},
if it can be represented in the form
$$
R^\simplex = \bigcap_{i=1}^m \left(R^{\simplex_i}
\times S^{\simplex\setminus\simplex_i}\right),
$$
where all $R^{\simplex_i}$ are proper consequences of $R^\simplex$.
For brevity we will omit the trivial multipliers in
intersections and write in the subsequent sections
expressions like $\bigcap_{i=1}^m R^{\simplex_i}$
instead of $\bigcap_{i=1}^m \left(R^{\simplex_i}\times
S^{\simplex\setminus\simplex_i}\right)$.

We see how to impose the structure of simplicial complex
on an amorphous set of points $X=\set{x_0, x_1,\ldots}$
via a relation $R^X$. The maximal simplices of $\Simplex$
must correspond to the irreducible components of the relation
 $R^X.$ Now we can evolve
--- starting only with a set of points and a relation on it
(in fact, we simply identify dimensions of the relation with the points) ---
the standard tools of the algebraic topology like homology,
cohomology, etc.

We wrote a program in \emph{\textbf{C}}
implementing the above constructions and manipulations
with them. Below we illustrate application of
the program to analysis of Conway's Game of Life
\cite{Gardner2}
and some of the Wolfram's elementary cellular automata
\cite{Wolfram}.

A few words are needed about computer implementation
of relations. To specify a $k$-ary relation $R^k$ we should
mark its points within the $k$-dimensional hypercube $S^k$,
i.e., define a \emph{characteristic function}
$\chi: S^k\rightarrow \set{0,1},$ with $\chi(\textbf{s})=1$ or 0
according as $\textbf{s}\in R^k$ or $\textbf{s}\notin R^k$.
Here $\textbf{s} = \left(s_0,s_1,\ldots,s_{k-1}\right)$
is a point of the hypercube. The simplest way to implement
the characteristic function is to enumerate all
the $q^k$ hypercube points in some standard,
e.g., lexicographic order:
\begin{center}
\begin{tabular}{ccccc|c}
$s_0$&$s_1$&$\ldots$&$s_{k-2}$&$s_{k-1}$&$i_{ord}$
\\
\hline
$0$&$0$&$\ldots$&
$0$&$0$& 0
\\
$1$&$0$&$\ldots$&
$0$&$0$& 1
\\[-5pt]
$\ \vdots$&$\ \vdots$&$\cdots$&$\ \vdots$&
$\ \vdots$&$\vdots$
\\
$q-2$&$~q-1$&$\ldots$&
$q-1$&$~q-1$&
$~q^k-2$
\\
$q-1$&$~q-1$&$\ldots$&
$q-1$&$~q-1$&$~q^k-1$
\\
\end{tabular}
\end{center}
Then the relation can be represented by a string of $q^k$ bits. We
call this string \emph{bit table} of relation. Symbolically
$\mathrm{BitTable}\left[\,i_{ord}\right]:= \left(\mathbf{s}\in
R^k\right).$ Note that $\mathbf{s}$ is a (``little-endian'')
representation of the number $i_{ord}$ in the base $q$. Most
manipulations with relations are reduced to very efficient bitwise
computer commands. Of course, symmetric or sparse (or, vice versa,
dense) relations can be represented in a more economical way, but
these are technical details of implementation.

\section{Conway's Game of Life}

The local rule of the cellular automaton \emph{\textbf{Life}}
is defined over the $10$-set $\simplex=\set{x_0,\ldots,x_9}$:
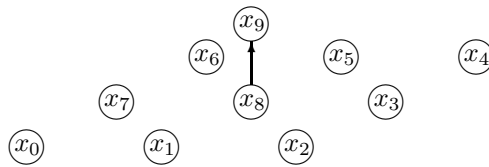
\begin{center}
{\setlength{\unitlength}{0.85pt}
\begin{picture}(230,45)(0,0)
\putxi{15}{0}{0}
\putxi{75}{0}{1}
\putxi{135}{0}{2}
\putxi{55}{20}{7}
\putxi{115}{20}{8}
\put(107.5,27.9){\vector(0,1){19}}
\putxi{115}{54.9}{9}
\putxi{175}{20}{3}
\putxi{95}{40}{6}
\putxi{155}{40}{5}
\putxi{215}{40}{4}
\end{picture}
}
\end{center}
Here the point $x_9$ is the next time step
of the point $x_8$. The state set $S$ is $\set{0,1}$.
The local rule can be represented as a relation
$\RLife$ on the 10-dimensional hypercube $S^\simplex.$
By definition, the hypercube element
belongs to the relation of the automaton \emph{\textbf{Life}},~ i.e.,~
 $\left(x_0,\ldots,x_9\right)\in\RLife$,~
in the following cases:
\begin{enumerate}
    \item
$\left(\sum_{i=0}^7x_i = 3\right) \wedge \left(x_9 = 1\right)$,
    \item
$\left(\sum_{i=0}^7x_i = 2\right) \wedge \left(x_8 = x_9\right)$,
    \item
$x_9 = 0$, if none of the above conditions holds.
\end{enumerate}

The number of elements of $\RLife$ is  $\left|\RLife\right|=512$.
The relation $\RLife$, as is the case
for any cellular automaton, is \emph{functional}:
the state of $x_9$ is uniquely determined by the states of other points.
The state set $S=\set{0,1}$ can be \emph{additionally}
endowed with the structure of the field $\F_2.$
We accompany the below analysis of the structure of
$\RLife$ by description in terms of
polynomials from
 $\F_2\left[x_0,\ldots,x_9\right].$
This is done only for illustrative purposes and for comparison
with the Gr\"obner basis method. In fact, we transform the
relations to polynomials only for output. This is done by
computationally very cheap Lagrange interpolation generalized to
the multivariate case. In the case $q=2$, the polynomial which set
of zeros corresponds to a relation is constructed uniquely. If
$q=p^n>2$, there is a freedom in the choice of nonzero values of
constructed polynomial, and the same relation can be represented
by many polynomials.

The polynomial representing $\RLife$ takes the form
\begin{equation}
\PLife= x_9
+x_8
\left\{
\sigma_7
+\sigma_6
+\sigma_3
+\sigma_2
\right\}
+\sigma_7
+\sigma_3,
\label{polylife}
\end{equation}
where $\sigma_k\equiv\esymm{k}{x_0,\ldots,x_7}$ is
the $k$th \emph{elementary symmetric polynomial} defined
for  $n$ variables
$x_0,\ldots,x_{n-1}$ by the formula:
$$
\esymm{k}{x_0,\ldots,x_{n-1}} =
\sum\limits_{0\leq i_0<i_1<\cdots<i_{k-1}<n}x_{i_0}x_{i_1}\cdots x_{i_{k-1}}.
$$

The relation $\RLife$
is reducible. It decomposes into
two equivalence classes (with respect to
the permutations of the points $x_0,\ldots,x_7$)
of relations defined over 9 points:
\begin{enumerate}
    \item \label{class1}
Eight relations $R_1^{\simplex\setminus\{x_i\}},~~~
0\leq i\leq7.$

Their polynomials $P_1^{i}\left(x_0,\ldots,\widehat{x_i},\ldots,
x_7,x_8,x_9\right)$
take the form
$$
P_1^i=
x_8x_9
\left\{
\sigma^i_6
+\sigma^i_5
+\sigma^i_2
+\sigma^i_1
\right\}
+x_9
\left\{
\sigma^i_6
+\sigma^i_2
+1
\right\}
+x_8
\left\{
\sigma^i_7
+\sigma^i_6
+\sigma^i_3
+\sigma^i_2
\right\},
$$
\begin{equation}
\sigma^i_k\equiv\esymm{k}{x_0,\ldots,
\widehat{x_i},\ldots,x_7}.
\label{poly1}
\end{equation}
    \item  \label{class2}
One relation $R_2^{\simplex\setminus\{x_8\}}$
with polynomial
$P_2^{8}\left(x_0,\ldots, x_7,x_9\right)$:
\begin{equation}
P_2^{8}=x_9
\left\{
\sigma_7
+\sigma_6
+\sigma_3
+\sigma_2
+1
\right\}
+\sigma_7
+\sigma_3,~~~~\sigma_k\equiv\esymm{k}{x_0,\ldots,x_7}.
\label{poly2}
\end{equation}
\end{enumerate}
The relation $\RLife$ has the following decomposition
\begin{equation}
\RLife =
R_2^{\simplex\setminus\{x_8\}}\bigcap
\left(\bigcap\limits_{k=0}^6
R_1^{\simplex\setminus\{x_{i_k}\}}\right),
\label{lifedecomp}
\end{equation}
where $(i_0,\ldots,i_6)$ are any 7 different indices from the set
$(0,\ldots,7)$.

We see that the rule of \emph{\textbf{Life}} is defined on 8-dimensional
space-time simplices. Of course, this interpretation is based
on the concepts of the abstract combinatorial topology and differs
from the native interpretation of the game of \emph{\textbf{Life}}
as a (2+1)-dimensional lattice structure.

The relations $R_1^{\simplex\setminus\{x_i\}}$ and
$R_2^{\simplex\setminus\{x_8\}}$ are irreducible but not
prime, i.e., they have proper consequences.

The relation $R_1^{\simplex\setminus\{x_i\}}$ has two
classes of 7-dimensional
consequences:
\begin{enumerate}
    \item
Seven
 relations $R_{1.1}^{\simplex\setminus\{x_i,x_j\}}$ with polynomials
\begin{eqnarray}
P_{1.1}^{ij}
\left(x_0,\ldots,\widehat{x_i},\ldots,
\widehat{x_j},\ldots,x_7,x_8,x_9\right)=
~~~~~~~~~~~~~~~~~~~~~~~~~~~~~~~~
&&
\nonumber
\\
x_8x_9
\left\{
\sigma^{ij}_6
+\sigma^{ij}_5
+\sigma^{ij}_4
+\sigma^{ij}_3
+\sigma^{ij}_2
+\sigma^{ij}_1
+1
\right\}
&&
\nonumber
\\
+x_9
\left\{
\sigma^{ij}_6
+\sigma^{ij}_5
+\sigma^{ij}_3
+\sigma^{ij}_2
+\sigma^{ij}_1
+1
\right\},
&&
\label{poly11}
\\[5pt]
\sigma^{ij}_k\equiv
\esymm{k}{x_0,\ldots,\widehat{x_i},\ldots,\widehat{x_j},\ldots,x_7}.
\nonumber
\end{eqnarray}
    \item
One relation $R_{1.2}^{\simplex\setminus\{x_i,x_8\}}$ with
polynomial
\begin{equation}
P_{1.2}^{i}\left(x_0,\ldots,\widehat{x_i},\ldots,x_7,x_9\right)=
x_9
\left\{
\sigma^{i}_7
+\sigma^{i}_6
+\sigma^{i}_5
+\sigma^{i}_3
+\sigma^{i}_2
+\sigma^{i}_1
+1
\right\}.
\label{poly12}
\end{equation}
\end{enumerate}
The 8-dimensional relation $R_2^{\simplex\setminus\{x_8\}}$ has one
class of  7-dimensional consequences.
This class contains 8 already obtained relations
$R_{1.2}^{\simplex\setminus\{x_i,x_8\}}$
with polynomials (\ref{poly12}).

Continuing the process of construction of decompositions
and proper consequences we come finally to the prime
relations $R^{\simplex_{i_0i_1i_2i_3}}$ defined over
4-simplices
$\simplex_{i_0i_1i_2i_3}=
\set{x_{i_0},x_{i_1},x_{i_2},x_{i_3},x_9}$,
where $i_k\in \set{0,1,\ldots,7}$ and $i_0<i_1<i_2<i_3$.
The polynomials of these relations take the form
\begin{equation}
P^{i_0,i_1,i_2,i_3} = x_9\esymm{4}{x_{i_0},x_{i_1},x_{i_2},x_{i_3}}
\equiv x_9x_{i_0}x_{i_1}x_{i_2}x_{i_3}.
\label{poly0123}
\end{equation}

Substituting (\ref{poly0123}) in (\ref{poly1}), (\ref{poly2}),
(\ref{poly11}), and (\ref{poly12}) (this is a \emph{purely
polynomial} simplification) we have finally the following
polynomial form of the system of relations valid for the
\emph{\textbf{Life}} rule:
\begin{eqnarray}
x_8x_9
\left\{
\sigma^i_2
+\sigma^i_1
\right\}
+x_9
\left\{
\sigma^i_2
+1
\right\}
+x_8
\left\{
\sigma^i_7
+\sigma^i_6
+\sigma^i_3
+\sigma^i_2
\right\}&=&0,
\label{poly1red}
\\
x_9
\left\{
\sigma_3
+\sigma_2
+1
\right\}
+\sigma_7
+\sigma_3&=&0,
\label{poly2red}
\\
\left(x_8x_9
+x_9\right)
\left\{
\sigma^{ij}_3
+\sigma^{ij}_2
+\sigma^{ij}_1
+1
\right\}&=&0,
\label{poly11red}
\\
x_9
\left\{
\sigma^{i}_3
+\sigma^{i}_2
+\sigma^{i}_1
+1
\right\}&=&0,
\label{poly12red}
\\
x_9x_{i_0}x_{i_1}x_{i_2}x_{i_3}&=&0.
\label{poly0123red}
\end{eqnarray}
Relations (\ref{poly0123red}) have a simple interpretation:
if the point $x_9$ is in the state 1, then
at least one of any four points surrounding the center
$x_8$ must be in the state 0.

The above analysis of the relation $\RLife$
takes
$< 1$ sec on a 1.8GHz AMD Athlon notebook with 960Mb. %under the Windows XP.

To compute the Gr\"obner basis we must add to polynomial
 (\ref{polylife}) ten polynomials
\begin{equation}
 x_i^2+x_i,\ i = 0,\ldots,9,
\label{Fermat2}
\end{equation}
expressing the relation $x^{p^n}=x$ valid
for all elements of any finite field $\F_{p^n}$.

Computation of the Gr\"obner basis over $\F_2$ with the
help of Maple 9 gives the following. Computation for the
pure lexicographic order with the variable ordering
$x_9\succ x_8\succ\cdots\succ x_0$ remains initial polynomial
(\ref{polylife}) unchanged, i.e., does not give any
additional information. The pure lexicographic order
with the variable ordering  $x_0\succ x_1\succ\cdots\succ x_9$ gives
relations (\ref{poly1red})---(\ref{poly0123red}) (modulo
several polynomial reductions violating the symmetry
of polynomials). The computation takes 1 h 22 min.
%21 min 37 sec.
Computation for the degree-reverse-lexicographic
order also gives relations
(\ref{poly1red})---(\ref{poly0123red}) (with the above
reservation). The times are 51 min
%50 min 47 sec
for the variable ordering $x_0\succ x_1\succ\cdots\succ x_9$, and 33 min
%17 sec
for the ordering $x_9\succ x_8\succ\ldots\succ x_0$.

\section{Elementary Cellular Automata}

Simplest binary, nearest-neighbor, one-dimensional
cellular automata were called \emph{elementary cellular automata}
by S. Wolfram, who has extensively studied their properties
\cite{Wolfram}. A large collection of results concerning these automata
is presented in the Wolfram's online atlas \cite{site}.
In the exposition below we use Wolfram's notations and terminology.
The elementary cellular automata
are simpler than the \emph{\textbf{Life}}, and we
may pay more attention to the topological aspects
of our approach.

Local rules of the elementary cellular automata are defined
on the 4-set
$\simplex=\set{p,q,r,s}$ which can be pictured by the icon
\Wpqrsbare{0}{0}.
A local rule is a binary function of the form $s=f(p,q,r).$
There are totally $2^{2^3}=256$ local rules,
each of which can be indexed with an 8-bit binary number.

Our computation with relations representing the local
rules shows that the total number 256 of them is divided
into 118 reducible and 138 irreducible relations.
Only two of the irreducible relations appeared to be prime,
namely, the rules 105 and 150%
\footnote{They are represented by the linear polynomial
equations $p+q+r+s+1=0$ and  $p+q+r+s=0$  for the rules
105 and 150, respectively.} in Wolfram's numeration.%
\footnote{Wolfram prefers ``big-endian'' representation of binary numbers.}

We consider the elementary automata on a space-time
lattice with integer coordinates
$(x,t)$, i.e., $x\in \Z =\set{\ldots,-1,0,1,\ldots}$ or
$x\in \Z_m$ (spatial $m$-periodicity),
$~ t\in \Z^*=\set{0,1,\ldots}.$
We denote a state of the
point on the lattice by $u(x,t)\in S=\set{0,1}$.
Generally the points  are connected
as is shown on the $5\times3$ fragment of the lattice
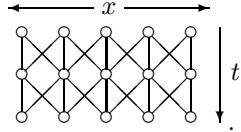
\begin{center}
\begin{picture}(78,45)(0,-5)
\put(27,41){\vector(-1,0){32}}
\put(33,41){\makebox(0,0){$x$}}
\put(39,41){\vector(1,0){32}}
\put(75,34){\vector(0,-1){36}}
\put(81,17){\makebox(0,0){$t$}}
\multiput(0,32)(16,0){5}{\circle{4}}
\multiput(1.4,30.4)(16,0){4}{\line(1,-1){13}}
\multiput(0,18)(16,0){5}{\line(0,1){12}}
\multiput(1.4,17.4)(16,0){4}{\line(1,1){13}}
\multiput(0,16)(16,0){5}{\circle{4}}
\multiput(1.4,14.4)(16,0){4}{\line(1,-1){13}}
\multiput(1.4,1.4)(16,0){4}{\line(1,1){13}}
\multiput(0,2)(16,0){5}{\line(0,1){12}}
\multiput(0,0)(16,0){5}{\circle{4}}
\end{picture}.
\end{center}
There are no horizontal ties due to the fundamental
property of cellular automata --- the states of points
at a given temporal layer are independent.

Applying our approach we see that some automata with
reducible local relations can be decomposed into automata
on disjoint unions of subcomplexes:
\begin{enumerate}
    \item
Two automata 0 and 255 are defined on
disjoint union of vertices.
    \item
Six automata 15, 51, 85, 170, 204 and 240 are, in fact,
disjoint collections of zero-dimensional automata.
What we call \emph{zero-dimensional automaton}
is spatially zero-dimensional analog
of the Wolfram's elementary automaton, i.e.,
a single cell evolving with time. There are, obviously,
four such automata with local relations represented by the bit tables
\begin{eqnarray}
&&1100,\nonumber\\
&&0110,
\label{oscpoint}
\\
&&1001,\nonumber\\
&&0011.\nonumber
\end{eqnarray}
We call the automaton with bit table
(\ref{oscpoint}) \emph{oscillating point}
since its time evolution consists in periodic changing 0
by 1 and vice versa. It is easy to ``integrate'' these automata.
Their general solutions are respectively
\begin{eqnarray}
u(t)&=&0,\nonumber\\
u(t)&=&u(0)+t\mod 2,~~~
\label{oscsol}
\text{\emph{oscillating point,}}\\
u(t)&=&u(0),\nonumber\\
u(t)&=&1.\nonumber
\end{eqnarray}

As an example consider the rule 15.
The local relation is defined on the set
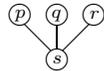
\Wpqrsbare{0}{0} and its bit table  is 0101010110101010.
This relation is reduced to the relation on the face
\Wpsbare{0}{0} and its bit table 0110 coincides with
bit table (\ref{oscpoint}) of the oscillating point.
We see that the automaton 15 decomposes into the union
of identical zero-dimensional automata on the disconnected lattice
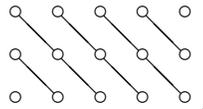
\begin{center}
\begin{picture}(70,45)(0,-5)
\multiput(0,32)(16,0){5}{\circle{4}}
\multiput(1.4,30.4)(16,0){4}{\line(1,-1){13}}
\multiput(0,16)(16,0){5}{\circle{4}}
\multiput(1.4,14.4)(16,0){4}{\line(1,-1){13}}
\multiput(0,0)(16,0){5}{\circle{4}}
\end{picture}.
\end{center}
Using (\ref{oscsol}) we can write the general solution for the automaton 15
$$
u(x,t) = u(x-t,0)+t\mod 2.
$$
    \item
Ten automata 5, 10, 80, 90, 95, 160, 165, 175,
245, 250 are decomposed into two identical  automata.

As an example let us consider the rule 90.
This automaton is distinguished as producing the fractal
(of the topological dimension 1 and Hausdorff dimension
$\ln3/\ln2\approx1.58$) known
as the Sierpinski sieve, Sierpinski gasket, or Sierpinski triangle.
Its local relation on the set \Wpqrsbare{0}{0}
is represented by the bit table 1010010101011010.
The relation is reduced to the relation on the face
\Wprsbare{0}{0} with the bit table
\begin{equation}
10010110.
\label{bt90}
\end{equation}

From the structure of the domain of the reduced relation it is
clear that the lattice decomposes into two identical independent
lattices as is shown
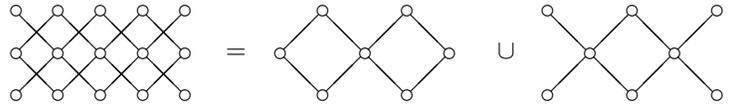
\begin{center}
\begin{picture}(70,25)(0,14)
\multiput(0,32)(16,0){5}{\circle{4}}
\multiput(1.4,30.4)(16,0){4}{\line(1,-1){13}}
\multiput(1.4,17.4)(16,0){4}{\line(1,1){13}}
\multiput(0,16)(16,0){5}{\circle{4}}
\multiput(1.4,14.4)(16,0){4}{\line(1,-1){13}}
\multiput(1.4,1.4)(16,0){4}{\line(1,1){13}}
\multiput(0,0)(16,0){5}{\circle{4}}
\end{picture}$~~=~~$
\begin{picture}(70,25)(0,14)
\multiput(16,32)(32,0){2}{\circle{4}}
\multiput(17.4,30.4)(32,0){2}{\line(1,-1){13}}
\multiput(1.4,17.4)(32,0){2}{\line(1,1){13}}
\multiput(0,16)(32,0){3}{\circle{4}}
\multiput(1.4,14.4)(32,0){2}{\line(1,-1){13}}
\multiput(17.4,1.4)(32,0){2}{\line(1,1){13}}
\multiput(16,0)(32,0){2}{\circle{4}}
\end{picture}
$~~\cup~~$
\begin{picture}(70,25)(0,14)
\multiput(0,32)(32,0){3}{\circle{4}}
\multiput(1.4,30.4)(32,0){2}{\line(1,-1){13}}
\multiput(17.4,17.4)(32,0){2}{\line(1,1){13}}
\multiput(16,16)(32,0){2}{\circle{4}}
\multiput(17.4,14.4)(32,0){2}{\line(1,-1){13}}
\multiput(1.4,1.4)(32,0){2}{\line(1,1){13}}
\multiput(0,0)(32,0){3}{\circle{4}}
\end{picture}.
\end{center}
\vspace*{10pt}
To find a general solution of the automaton 90 it
is convenient to transform bit table (\ref{bt90}) to
an algebraic relation. It is the linear relation
$s+p+r=0$
and the general solution of the automaton takes the form
$$
u(x,t)=\sum\limits_{k=0}^t \binom{t}{k}u(x-t+2k,0)\mod 2.
$$
\end{enumerate}

In the above examples we have considered the automata with
reducible relations. If a local relation is irreducible but
has proper consequences we also, in some cases, can
obtain a useful information.

For example, there are 64 automata%
\footnote{The full list of these automata in the Wolfram's numeration
is
  2,   4,   8,  10,  16,  32,  34,  40,
 42,  48,  64,  72,  76,  80,  96, 112,
128, 130, 132, 136, 138, 140, 144, 160,
162, 168, 171, 174--176, 186, 187,
190--192, 196, 200, 205, 206, 208,
220, 222--224, 234--239,
241--254.}
--- both reducible and irreducible ---
having proper consequencies with the
bit table
\begin{equation}
1101
\label{finiterod}
\end{equation}
on one or two or three of the following faces
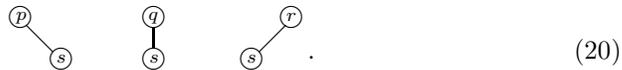
\begin{equation}
\text{\Wpsbare{0}{0}~~~~~~ \Wqsbare{0}{0}~~~~~~
\Wrsbare{0}{0}.}
\label{finiterodsets}
\end{equation}
The algebraic forms of relation
(\ref{finiterod}) on faces (\ref{finiterodsets}) are
$$
ps+s=0,~~~qs+s=0,~~~rs+s=0,
$$ respectively.

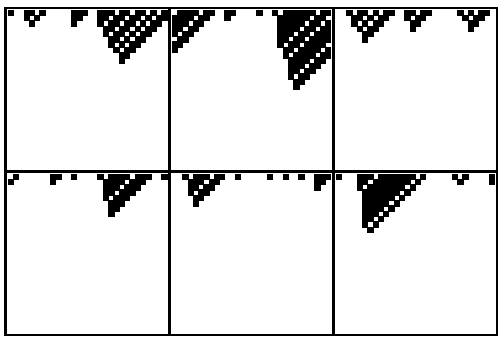
\begin{figure}
\begin{center}
\begin{picture}(188,126)(0,0)
\put(1,1){\line(1,0){186}}
\put(1,63){\line(1,0){186}}
\put(1,125){\line(1,0){186}}
\put(1,1){\line(0,1){124}}
\put(63,1){\line(0,1){124}}
\put(125,1){\line(0,1){124}}
\put(187,1){\line(0,1){124}}
%====== 1 ======
%------ 1.1 ------
\put(3,123){\blcell}
\multiput(9,123)(2,0){2}{\blcell}
\put(15,123){\blcell}
\multiput(27,123)(2,0){3}{\blcell}
\multiput(37,123)(2,0){3}{\blcell}
\multiput(45,123)(2,0){2}{\blcell}
\multiput(51,123)(2,0){2}{\blcell}
\put(57,123){\blcell}
\put(61,123){\blcell}
%------ 1.2 ------
\put(9,121){\blcell}
\put(13,121){\blcell}
\multiput(27,121)(2,0){2}{\blcell}
\multiput(37,121)(2,0){2}{\blcell}
\multiput(43,121)(2,0){2}{\blcell}
\multiput(49,121)(2,0){2}{\blcell}
\put(55,121){\blcell}
\multiput(59,121)(2,0){2}{\blcell}
%------ 1.3 ------
\put(11,119){\blcell}
\put(27,119){\blcell}
\put(37,119){\blcell}
\multiput(41,119)(2,0){2}{\blcell}
\multiput(47,119)(2,0){2}{\blcell}
\put(53,119){\blcell}
\multiput(57,119)(2,0){2}{\blcell}
%------ 1.4 ------
\multiput(39,117)(2,0){2}{\blcell}
\multiput(45,117)(2,0){2}{\blcell}
\put(51,117){\blcell}
\multiput(55,117)(2,0){2}{\blcell}
%------ 1.5 ------
\put(39,115){\blcell}
\multiput(43,115)(2,0){2}{\blcell}
\put(49,115){\blcell}
\multiput(53,115)(2,0){2}{\blcell}
%------ 1.6 ------
\multiput(41,113)(2,0){2}{\blcell}
\put(47,113){\blcell}
\multiput(51,113)(2,0){2}{\blcell}
%------ 1.7 ------
\put(41,111){\blcell}
\put(45,111){\blcell}
\multiput(49,111)(2,0){2}{\blcell}
%------ 1.8 ------
\put(43,109){\blcell}
\multiput(47,109)(2,0){2}{\blcell}
%------ 1.9 ------
\multiput(45,107)(2,0){2}{\blcell}
%------ 1.10 ------
\put(45,105){\blcell}
%====== 2 ======
%------ 2.1 ------
\multiput(67,123)(2,0){4}{\blcell}
\multiput(77,123)(2,0){2}{\blcell}
\multiput(85,123)(2,0){2}{\blcell}
\put(97,123){\blcell}
\put(103,123){\blcell}
\multiput(107,123)(2,0){6}{\blcell}
\multiput(121,123)(2,0){2}{\blcell}
%------ 2.2 ------
\multiput(65,121)(2,0){4}{\blcell}
\multiput(75,121)(2,0){2}{\blcell}
\put(85,121){\blcell}
\multiput(105,121)(2,0){6}{\blcell}
\multiput(119,121)(2,0){2}{\blcell}
%------ 2.3 ------
\multiput(65,119)(2,0){3}{\blcell}
\multiput(73,119)(2,0){2}{\blcell}
\multiput(105,119)(2,0){5}{\blcell}
\multiput(117,119)(2,0){2}{\blcell}
\put(123,119){\blcell}
%------ 2.4 ------
\multiput(65,117)(2,0){2}{\blcell}
\multiput(71,117)(2,0){2}{\blcell}
\multiput(105,117)(2,0){4}{\blcell}
\multiput(115,117)(2,0){2}{\blcell}
\multiput(121,117)(2,0){2}{\blcell}
%------ 2.5 ------
\put(65,115){\blcell}
\multiput(69,115)(2,0){2}{\blcell}
\multiput(105,115)(2,0){3}{\blcell}
\multiput(113,115)(2,0){2}{\blcell}
\multiput(119,115)(2,0){3}{\blcell}
%------ 2.6 ------
\multiput(67,113)(2,0){2}{\blcell}
\multiput(105,113)(2,0){2}{\blcell}
\multiput(111,113)(2,0){2}{\blcell}
\multiput(117,113)(2,0){4}{\blcell}
%------ 2.7 ------
\multiput(65,111)(2,0){2}{\blcell}
\put(105,111){\blcell}
\multiput(109,111)(2,0){2}{\blcell}
\multiput(115,111)(2,0){4}{\blcell}
%------ 2.8 ------
\put(65,109){\blcell}
\multiput(107,109)(2,0){2}{\blcell}
\multiput(113,109)(2,0){4}{\blcell}
\put(123,109){\blcell}
%------ 2.9 ------
\put(107,107){\blcell}
\multiput(111,107)(2,0){4}{\blcell}
\multiput(121,107)(2,0){2}{\blcell}
%------ 2.10 ------
\multiput(109,105)(2,0){4}{\blcell}
\multiput(119,105)(2,0){2}{\blcell}
%------ 2.11 ------
\multiput(109,103)(2,0){3}{\blcell}
\multiput(117,103)(2,0){2}{\blcell}
%------ 2.12 ------
\multiput(109,101)(2,0){2}{\blcell}
\multiput(115,101)(2,0){2}{\blcell}
%------ 2.13 ------
\put(109,99){\blcell}
\multiput(113,99)(2,0){2}{\blcell}
%------ 2.14 ------
\multiput(111,97)(2,0){2}{\blcell}
%------ 2.15 ------
\put(111,95){\blcell}
%====== 3 ======
%------ 3.1 ------
\put(131,123){\blcell}
\multiput(135,123)(2,0){2}{\blcell}
\put(141,123){\blcell}
\multiput(145,123)(2,0){2}{\blcell}
\multiput(153,123)(2,0){2}{\blcell}
\multiput(159,123)(2,0){2}{\blcell}
\put(173,123){\blcell}
\put(177,123){\blcell}
\multiput(181,123)(2,0){2}{\blcell}
%------ 3.2 ------
\multiput(133,121)(2,0){2}{\blcell}
\put(139,121){\blcell}
\multiput(143,121)(2,0){2}{\blcell}
\put(153,121){\blcell}
\multiput(157,121)(2,0){2}{\blcell}
\put(175,121){\blcell}
\multiput(179,121)(2,0){2}{\blcell}
%------ 3.3 ------
\multiput(133,119)(4,0){2}{\blcell}
\multiput(141,119)(2,0){2}{\blcell}
\multiput(155,119)(2,0){2}{\blcell}
\multiput(177,119)(2,0){2}{\blcell}
%------ 3.4 ------
\put(135,117){\blcell}
\multiput(139,117)(2,0){2}{\blcell}
\put(155,117){\blcell}
\put(177,117){\blcell}
%------ 3.5 ------
\multiput(137,115)(2,0){2}{\blcell}
%------ 3.6 ------
\put(137,113){\blcell}
%====== 4 ======
%------ 4.1 ------
\put(5,61){\blcell}
\multiput(19,61)(2,0){2}{\blcell}
\put(27,61){\blcell}
\put(37,61){\blcell}
\multiput(41,61)(2,0){4}{\blcell}
\multiput(51,61)(2,0){3}{\blcell}
\put(61,61){\blcell}
%------ 4.2 ------
\put(3,59){\blcell}
\put(19,59){\blcell}
\multiput(39,59)(2,0){4}{\blcell}
\multiput(49,59)(2,0){3}{\blcell}
%------ 4.3 ------
\multiput(39,57)(2,0){3}{\blcell}
\multiput(47,57)(2,0){3}{\blcell}
%------ 4.4 ------
\multiput(39,55)(2,0){2}{\blcell}
\multiput(45,55)(2,0){3}{\blcell}
%------ 4.5 ------
\put(39,53){\blcell}
\multiput(43,53)(2,0){3}{\blcell}
%------ 4.6 ------
\multiput(41,51)(2,0){3}{\blcell}
%------ 4.7 ------
\multiput(41,49)(2,0){2}{\blcell}
%------ 4.8 ------
\put(41,47){\blcell}
%====== 5 ======
%------ 5.1 ------
\put(69,61){\blcell}
\multiput(73,61)(2,0){3}{\blcell}
\multiput(81,61)(2,0){2}{\blcell}
\put(89,61){\blcell}
\put(101,61){\blcell}
\put(107,61){\blcell}
\put(113,61){\blcell}
\multiput(119,61)(2,0){3}{\blcell}
%------ 5.2 ------
\multiput(71,59)(2,0){3}{\blcell}
\multiput(79,59)(2,0){2}{\blcell}
\multiput(119,59)(2,0){2}{\blcell}
%------ 5.3 ------
\multiput(71,57)(2,0){2}{\blcell}
\multiput(77,57)(2,0){2}{\blcell}
\put(119,57){\blcell}
%------ 5.4 ------
\put(71,55){\blcell}
\multiput(75,55)(2,0){2}{\blcell}
%------ 5.5 ------
\multiput(73,53)(2,0){2}{\blcell}
%------ 5.6 ------
\put(73,51){\blcell}
%====== 6 ======
%------ 6.1 ------
\put(127,61){\blcell}
\multiput(135,61)(2,0){3}{\blcell}
\multiput(143,61)(2,0){7}{\blcell}
\put(159,61){\blcell}
\put(171,61){\blcell}
\put(175,61){\blcell}
\multiput(185,61)(0,-2){2}{\blcell}
%------ 6.2 ------
\multiput(135,59)(2,0){2}{\blcell}
\multiput(141,59)(2,0){7}{\blcell}
\put(157,59){\blcell}
\put(173,59){\blcell}
%------ 6.3 ------
\put(135,57){\blcell}
\multiput(139,57)(2,0){7}{\blcell}
\put(155,57){\blcell}
%------ 6.4 ------
\multiput(137,55)(2,0){7}{\blcell}
\put(153,55){\blcell}
%------ 6.5 ------
\multiput(137,53)(2,0){6}{\blcell}
\put(151,53){\blcell}
%------ 6.6 ------
\multiput(137,51)(2,0){5}{\blcell}
\put(149,51){\blcell}
%------ 6.7 ------
\multiput(137,49)(2,0){4}{\blcell}
\put(147,49){\blcell}
%------ 6.8 ------
\multiput(137,47)(2,0){3}{\blcell}
\put(145,47){\blcell}
%------ 6.9 ------
\multiput(137,45)(2,0){2}{\blcell}
\put(143,45){\blcell}
%------ 6.10 ------
\put(137,43){\blcell}
\put(141,43){\blcell}
%------ 6.11 ------
\put(139,41){\blcell}
\end{picture}
\caption{Rule 168. Several random initial conditions\label{figu}}
\end{center}
\end{figure}

Relation (\ref{finiterod}) is non-functional.
Nevertheless, it imposes a severe restriction on the behavior
of the automata with such proper consequences. The peculiarities
in the behavior are clear visible in the atlas \cite{site},
where many results of computations with different initial
conditions are pictured. A typical pattern from this atlas
is reproduced in Fig. \ref{figu}, where several evolutions of the
automaton 168 are presented. The local relation of the automaton
168 is $pqr+qr+pr+s=0$. It has the proper consequence $rs+s=0.$
The black and white square cells in Fig. \ref{figu} correspond to 1's and
0's, respectively.
Note also that the authors of Fig. \ref{figu} have used
a spatially periodic condition. Their spacial variable is $x\in\Z_{30}.$

Relation (\ref{finiterod}) means that if,
say $r$, as for rule 168, is in the state 1 then $s$ may be
in both states 0 or 1, but if the state of $r$ is 0,
then the state of $s$ must be 0. Thus the corresponding diagonal
or vertical may contain either only 1's, or finite number
of initial 1's and then only 0's. The presence of a proper
consequence of the form (\ref{finiterod}) simplifies
essentially computation with such automata:
after the first appearance of 0, one can set 0's on all points along
the corresponding line.

In conclusion, let us present the results of analysis
of the automata 30 and 110. These automata are of special interest.
The automaton 30 demonstrates chaotic behavior
and even used as the random number generator in
\emph{\textbf{Mathematica.}} The automaton 110 is, like
a Turing machine, \emph{universal}, i.e., it is capable
of simulating any computational process, in particular,
any other cellular automaton.
The relations of both automata are irreducible but not prime.

The relation of automaton 30 is
$$
1001010101101010
$$
or in the algebraic form $$qr+s+r+q+p=0.$$
It has two proper consequences:

\begin{tabular}{lll}
face&\Wpqsbare{0}{0}
&
\Wprsbare{0}{0}\\[10pt]
bit table&11011110 & 11011110\\[10pt]
polynomial \hspace*{20pt}&$qs+pq+q$\hspace*{20pt} & $rs+pr+r$.\\[10pt]
\end{tabular}\\
The principal factor is $$1011111101111111 ~~~\text{or}~~~
qrs+pqr+rs+qs+pr+pq+s+p=0.$$
The Gr\"obner basis of automaton 30
in the total degree and reverse lexicographic
order is (omitting the trivial polynomials
 $p^2+p,~ q^2+q, ~r^2+r, ~s^2+s$)
$$\set{qr+s+r+q+p,~qs+pq+q,~rs+pr+r}.$$
We see that for the rule 30 the Gr\"obner basis polynomials coincide
with ours.

The relation of automaton 110 is
\begin{equation}
1100000100111110
\label{R110}
\end{equation}
or in the polynomial form $$pqr+qr+s+r+q=0.$$
The relation has three proper consequences:

\begin{tabular}{llll}
face&\Wpqsbare{0}{0}
&
\Wprsbare{0}{0}
&
\Wqrsbare{0}{0}
\\[10pt]
bit table&11011111 & 11011111 &10010111\\[10pt]
polynomial~~~&
$pqs+qs+pq+q~~~$ &
$prs+rs+pr+r~~~$ & $qrs+s+r+q$.\\[10pt]
\end{tabular}\\
The principal factor is
$$1111111111111110 ~~~\text{or}~~~ pqrs=0.$$
The Gr\"obner basis of automaton 110 contains different set of polynomials:
$$
\set{prs+rs+pr+r,~qs+rs+r+q,~qr+rs+s+q,~pr+pq+ps}.
$$
The system of relations defined by the
Gr\"obner basis
is:
\begin{eqnarray*}
R_1^{\set{p,r,s}}&=&11011111=\left\{prs+rs+pr+r=0\right\},
\\
R_2^{\set{q,r,s}}&=&10011111=\left\{qs+rs+r+q=0\right\},
\\
R_3^{\set{q,r,s}}&=&10110111=\left\{qr+rs+s+q=0\right\},
\\
R_4^{\set{p,q,r,s}}&=&1110101110111110=\left\{pr+pq+ps=0\right\}.
\end{eqnarray*}
\section{Conclusions}
Let us summarize the main novelties of the paper.
\begin{itemize}
    \item
We have introduced a notion of a
\emph{system of discrete relations on an abstract simplicial complex}.
Such a system can be interpreted as
\begin{itemize}
      \item
a natural generalization of the notion of cellular automaton;
      \item
a set-theoretic analog of a system of polynomial equations.
\end{itemize}
    \item
After introducing appropriate definitions,
we have developed and implemented algorithms for
\begin{itemize}
      \item
\emph{compatibility analysis} of a system of discrete relations;
    \item
constructing \emph{canonical decompositions} of discrete relations.
\end{itemize}
    \item
We have proposed a regular way to impose \emph{topology on
an arbitrary discrete relation} via its canonical decomposition:
identifying \emph{dimensions} of the relation with \emph{points}
and \emph{irreducible components} of the relation with
\emph{maximal simplices}, we define the structure of
an abstract simplicial complex on the relation under consideration.
    \item
Applying the above technique to some cellular automata
--- a special case of systems of discrete relations ---
we have obtained some new results. Most interesting of them,
in our opinion, is demonstration of how the presence
of non-trivial \emph{proper consequences} may determine
the global behavior of an automaton.
\end{itemize}
\section*{Acknowledgments}
This work was supported in part by the
grants
04-01-00784
from the Russian Foundation for Basic Research
 and
2339.2003.2
 from the Russian Ministry of Industry, Science and
Technologies.

\end{document}